\documentclass[%
 twocolumn,
groupedaddress,
 amsmath,amssymb,
 aps,
]{revtex4-2}

\usepackage{graphicx}
\usepackage{dcolumn}
\usepackage{bm}
\usepackage{xcolor}
\usepackage[flushleft]{threeparttable}
\newcolumntype{.}{D{.}{.}{-1}}
\newcolumntype{d}[1]{D{.}{.}{#1}}
\newcommand*{\wn}{cm$^{-1}$}
\newcommand*{\hsm}{H$_{2}$S}

\newcommand*{\X}{X$^1\Sigma_g^+$}

\newcommand*{\F}{F$^1\Sigma_g^+$}

\renewcommand{\eqref}[1]{Eq.~(\ref{#1})}

\begin{document}

\title{Shape resonances in H$_2$ as photolysis reaction intermediates}

\author{K.-F. Lai}
 \affiliation{Department of Physics and Astronomy, LaserLaB, Vrije Universiteit \\
 De Boelelaan 1081, 1081 HV Amsterdam, The Netherlands}
 \author{E. J. Salumbides}
 \affiliation{Department of Physics and Astronomy, LaserLaB, Vrije Universiteit \\
 De Boelelaan 1081, 1081 HV Amsterdam, The Netherlands}
 \author{W. Ubachs}%
  \affiliation{Department of Physics and Astronomy, LaserLaB, Vrije Universiteit \\
 De Boelelaan 1081, 1081 HV Amsterdam, The Netherlands}
 \author{M. Beyer}%
 \affiliation{Department of Physics and Astronomy, LaserLaB, Vrije Universiteit \\
 De Boelelaan 1081, 1081 HV Amsterdam, The Netherlands}

\date{\today}

\begin{abstract}
Shape resonances in H$_2$, produced as reaction intermediates in the photolysis of H$_2$S precursor molecules, are measured in a half-collision approach. Before desintegrating into two ground state H atoms, the reaction is quenched by two-photon Doppler-free excitation to the F electronically excited state of H$_2$. 
For $J=13,15,17,19$ and 21, resonances with lifetimes in the range of nano to milliseconds were observed with an accuracy of 
30~MHz (1.4~mK). 
The experimental resonance positions are found to be in excellent agreement with theoretical predictions when nonadiabatic and quantum electrodynamical corrections are included. This is the first time such effects are observed in collisions between neutral atoms.
From the potential energy curve of the H$_2$ molecule, now tested at high accuracy over a wide range of internuclear separations, the s-wave scattering length for singlet H(1s)+H(1s) scattering is determined at $a = 0.2735^{39}_{31}~a_0$.
\end{abstract}

\maketitle

The understanding of resonance phenomena occurring in the encounter of colliding particles is of eminent importance in physics. As was first discovered by Fermi and co-workers~\cite{Fermi1934}, particles colliding in a partial wave with non-zero rotational angular momentum can be trapped behind a potential barrier - a phenomenon known as {\it shape resonances} \cite{Breit1936}. These quasi-bound states are rare and occur incidentally through an interplay between the specific shape of a molecular potential and the centrifugal barrier $J(J+1)/2\mu R^2$, for a given partial wave $J$ and colliding particles with reduced mass $\mu$.
A second manifestation of collision resonances, called {\it Feshbach resonances}, have attracted great attention in the field of ultracold atoms in the last decade~\cite{Innouye1998,Theis2004,Chin2010}. 
Here the resonance occurs by coupling the continuum of one scattering channel to a bound molecular state from a different channel. This coupling can be the result of an intrinsic property of the molecule, like Coriolis interactions \cite{Glass-Maujean1978}, or it can be caused by external magnetic \cite{Regal2004} or electric \cite{Carrington1993b} fields.

Magnetic Feshbach resonances play an important role in the study of cold interacting Bose and Fermi gases, but experimental studies have been mainly limited to s-wave resonances. Recently, resonances for higher partial waves $J\ne 0$ have been observed~\cite{Yao2019,Gerken2019,Li2021}, which must be characterized as of mixed Feshbach/shape resonance character. A detailed understanding of higher partial wave magnetic Feshbach resonances requires therefore a good understanding of shape resonances~\cite{Gao2005}. 

Most Feshbach resonances have been observed using laser-coolable alkaline or earth-alkaline atoms, for which potential energy curves can be calculated to high accuracy, although in a phenomenological manner involving fitting of parameters~\cite{Jones2006,Hutson2006}.
The H+H collision system stands out representing the interaction between the simplest atomic constituents,
which allows the comparison of experimental results with full-fledged ab initio calculations.  The H(1s) + H(1s) scattering is of fundamental interest in physics, playing a role in the formation of molecular hydrogen in the universe \cite{Forrey2016}, frequency shifts in the hydrogen maser~\cite{Wittke1956}, in the precision metrology of atomic hydrogen and the determination of the Rydberg constant \cite{Matveev2019}, and for the formation of hydrogen Bose-Einstein condensates \cite{Fried1998}. The light mass of the hydrogen atom compared to all other studied systems, makes this collision uniquely sensitive to nonadiabatic effects, i.e., the effects of distant electronic states on a given collision channel \cite{Van-Vleck1936}.

\begin{figure}[b]
\begin{center}
\includegraphics[width=0.95\linewidth]{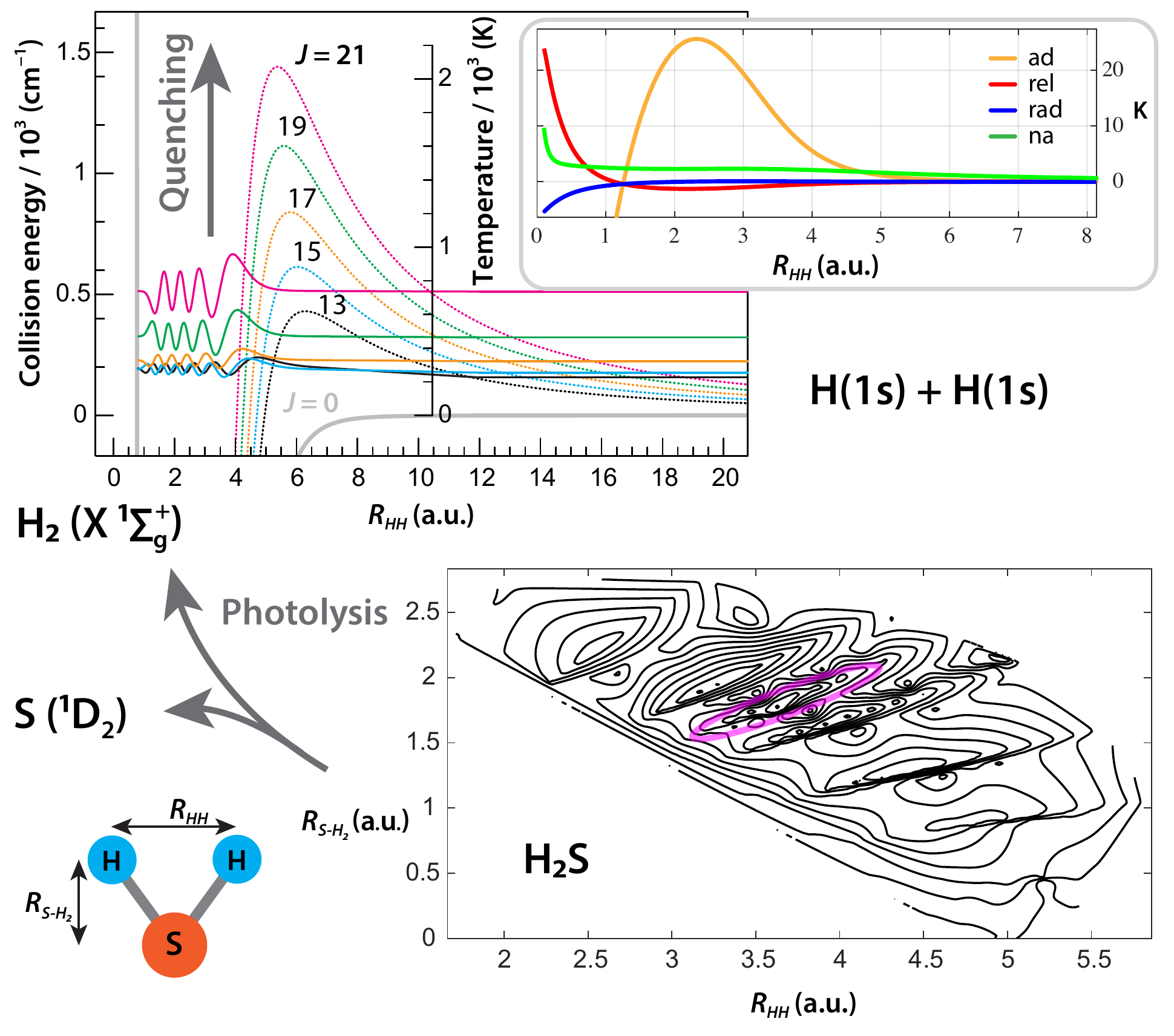}
\caption{\label{ex_scheme}
Potential energy surfaces for the electronic ground states of H$_2$S \cite{Tarczay2001} and H$_2$ \cite{Komasa2019}. The H$_2$S equilibrium geometry is indicated with a magenta ellipse in the contour plot. For H$_2$ the radial wave functions (not to scale) of the observed shape resonances and the respective rotational barriers are shown. The inset displays the $J$-independent contribution of beyond-Born-Oppenheimer contributions to the potential.
}
\end{center}
\end{figure}

We present in this letter the study of shape resonances in the collision of two ground state hydrogen atoms at an accuracy $10^{-5}$ relative to the collision energy. 
Five very narrow shape resonances with $J=13, 15, 17, 19$ and $21$, 
with predicted lifetimes varying from ns to ms~\cite{Selg2012},
have been observed at a precision high enough to identify for the first time nonadiabatic, as well as relativistic and QED effects in neutral atom-atom collisions. Previously nonadiabatic effects have been observed in the shape resonances of the ion-neutral H+H$^+$ collision at a hundred times lower precision \cite{Beyer2016,Beyer2018}. 

Being inaccessible to laser cooling with standard techniques, the study of controlled H+H collisions could take place in atomic beam scattering experiments or using a half-collision approach, i.e., by photodissociating molecular hydrogen. Given the narrow width of the resonances (50~nK--5~mK), the former seems unrealistic, whereas for the later only collisions between ground and excited H atoms could be probed~\cite{Meng2016}.

Instead we choose to photolyse hydrogen sulfide molecules at 281.8~nm, providing sufficient energy for a complete three particle dissociation~\cite{Zhou2020,Zhao2021}, according to:
\begin{equation}
    \text{H$_{2}$S} \xrightarrow{2h\nu} \text{S} (^{1}{\rm D}_{2}) + \text{H$_{2}^{*}$}(X^1\Sigma_g^+)
   \rightarrow \text{S} (^{1}{\rm D}_{2}) + \text{H(1s)} + \text{H(1s)}, \nonumber
\end{equation}
to produce quasi-bound H$_2^*$ molecules, existing as an intermediate reaction product before desintegrating into two H(1s) atoms. The reaction is interrupted - quenched - by probing the long-lived shape resonances H$_2^*$ via 
two-photon Doppler-free excitation into the  \F, $v = 0$ electronically excited outer well of the EF state (denoted as F0 in the following). A third UV laser pulse probes the F0 population by photoionization and the resulting H$_2^+$ ions are detected selectively using a multichannel plate detector after passing a time-of-flight mass separator. The latter distinguishes the weak H$_2^+$ signals from rather strong  H$_2$S$^+$, SH$^+$, S$^+$ and H$^+$ background signals. A more detailed description of the experimental setup and the narrow-band pulsed dye amplifier (PDA) laser is given in Ref.~\cite{Lai2021}. Calibrating the laser light using a wavemeter and accounting for frequency chirp induced by the amplifier, allows to reach an absolute accuracy of a few tens of MHz, corresponding to around 1~mK.

The \hsm~\cite{Tarczay2001} and H$_2$ \cite{Komasa2019} potential energy surfaces, displayed in Fig.~\ref{ex_scheme}, illustrate the excitation of the shape resonances. The mean distance between the protons in the equilibrium geometry of \hsm~(indicated by a magenta ellipse) is comparable to the outer turning point of the H$_2$ potential, where the H$_2^*$ radial wave functions have their largest amplitude, allowing an efficient production of the resonances in the spirit of the Franck-Condon principle. The radial wavefunctions of the shape resonances are depicted together with the respective centrifugal barriers, necessary for their formation. Centrifugal barriers with a height of a few hundred up to two thousand Kelvin are found for $J=15-21$, resulting in a strongly reduced tunneling probability and a tiny amplitude of the radial wave function for large internuclear distance $R$ (behind the barrier). The contribution of the leading terms that go beyond the Born-Oppenheimer approximation - adiabatic, relativistic, radiative and nonadiabatic corrections - are depicted in the inset.

\begin{figure}
\begin{center}
\includegraphics[width=1.0\linewidth]{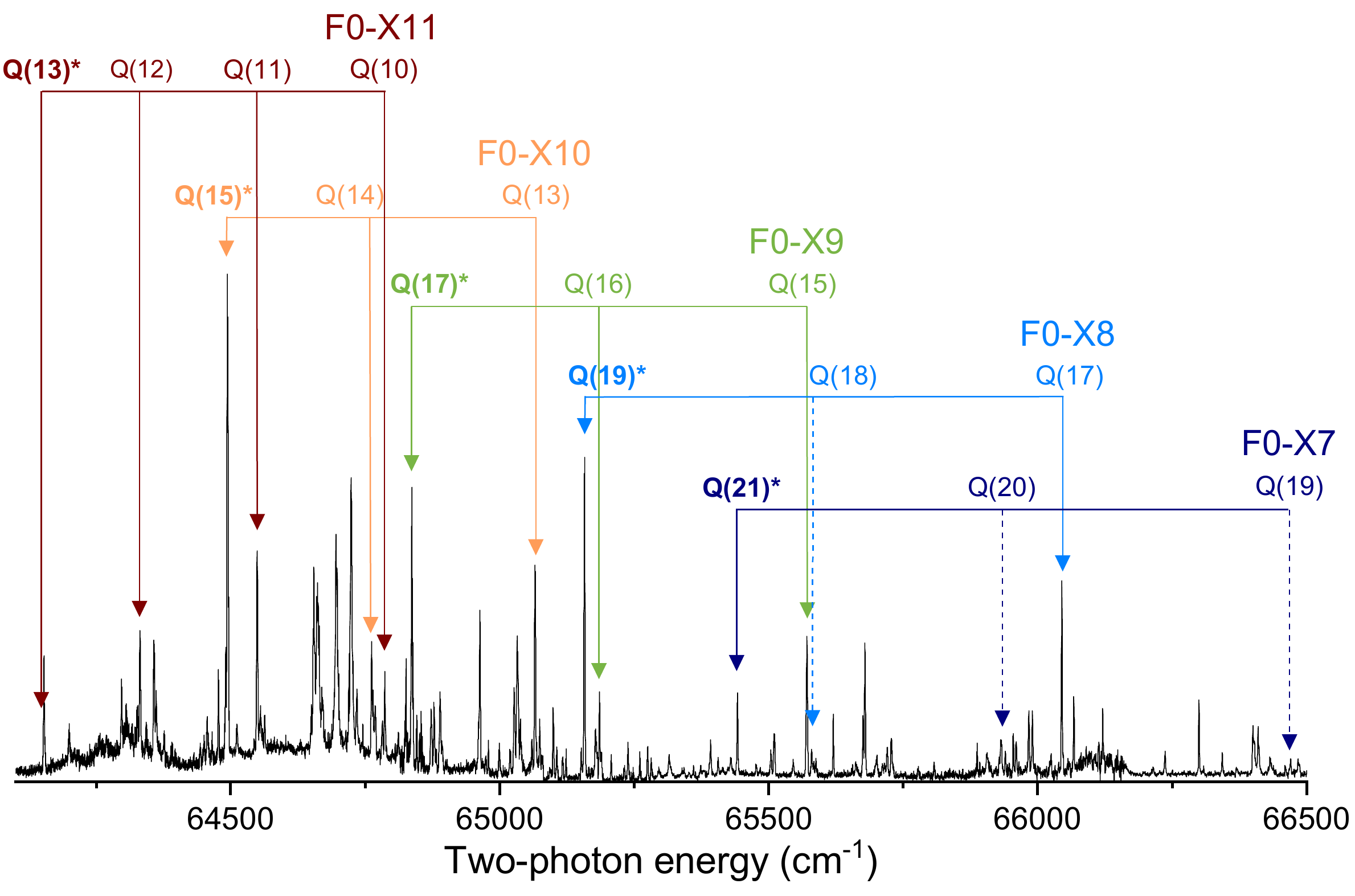}
\caption{\label{Overview}
Overview spectra recorded in a two-laser scheme with two-photon UV-photolysis of H$_2$S, followed by 2+1 REMPI on F0-X($v''$) bands with a UV-tunable frequency-doubled dye laser. Transitions are labeled with quantum numbers of the ground level.
Excitations from quasi-bound resonances are indicated with asterisk (*). 
}
\end{center}
\end{figure}

The product distribution of rovibrationally excited states H$_{2}^{*}$ was measured in a two-color overview scan in the wavelength range 300-310 nm (plotted in Fig.~\ref{Overview})
showing a forest of spectral lines in ($v',v''$) vibrational bands of the \F-\X\  system, detected via 2+1 resonance-enhanced multiphoton ionization.
In this dense overview spectrum progressions of O($\Delta J=-2$), Q($\Delta J=0$) and S($\Delta J=2$) rotational branches in the various bands are detected.
The assignment of the bound states in Fig.~\ref{Overview}, and in particular of the shape resonances is based on 
a comparison with calculated values of the transition frequencies. 

For this purpose highly accurate calculations of X($v,J$) levels were performed using the nonrelativistic quantum electrodynamics (NRQED) approach, in which relativistic, leading-order radiative and higher order QED corrections are added to a nonrelativistic Hamiltonian:
\begin{equation}
E(\alpha)= m_\text{e}\alpha^{2}E^{(2)}+m_\text{e}\alpha^{4}E^{(4)} + m_\text{e}\alpha^{5}E^{(5)}\cdots.
\end{equation}
The nonrelativistic energies $E^{(2)}$ are evaluated using non-adiabatic perturbation theory (NAPT) \cite{Czachorowski2018} accounting for terms up to $m_\text{e}\alpha^2(m_\text{e}/m_\text{p})^2$, while maintaining the computational efficiency of the Born-Oppenheimer (BO) approach through separating the electronic and nuclear Schr\"odinger equation. This is achieved by including $R$-dependent corrections to the potential energy curve and by employing $R$-dependent reduced masses in the nuclear Schr\"odinger equation  $\mu_{\|}(R)$ and $\mu_{\perp}(R)$ defined in~\cite{Komasa2019}.
The radial nuclear Schr\"odinger equation (in atomic units) within NAPT is given by \cite{Komasa2019}
\begin{align}\label{eq:SEraw}
    \left[-\frac{1}{R^{2}} \frac{\partial}{\partial R} \frac{R^{2}}{2 \mu_{\|}(R)} \frac{\partial}{\partial R}+\frac{J(J+1)}{2 \mu_{\perp}(R) R^{2}}+\mathcal{V}(R)\right] \phi_{i}(R) \notag\\
=E_i \phi_{i}(R).
\end{align}
Applying the ansatz
$\chi_i(R)=R\phi_i(R)\exp{(-Z(R))}$ to remove the first-order radial derivative~\cite{Leroy1985,Selg2011}, the Schr\"odinger equation can be numerically solved using $\mathcal{V}(R) = \mathcal{E}_\text{BO} + \mathcal{E}_\text{ad} + \delta\mathcal{E}_\text{na}$, being the BO \cite{Pachucki2010}, adiabatic \cite{Pachucki2014} and non-adiabatic \cite{Pachucki2015} potential energy curve, respectively.
Relativistic ($m_\text{e}\alpha^4$) and QED ($m_\text{e}\alpha^5$, $m\alpha^6$) corrections are included in the computation as described in Ref.~\cite{Komasa2019}. The inset in Fig.~\ref{ex_scheme} displays these corrections and illustrates their size compared to the BO energies.

We add an extension to the NAPT framework to calculate the positions and widths of the shape resonances using the time-delay matrix technique \cite{Smith1960a}, with the phase shift $\eta_{J}$ obtained for a given $J$ by propagating the wave function to large internuclear distance, where $\lim_{R\to\infty} Z(R) = 0$, so that standard scattering boundary conditions can be applied.

The present numerical calculation reaches an accuracy on the order of 0.003~\wn\ for bound states and resonances, with the uncertainty originating mainly from terms proportional to $m_\text{e}\alpha^2(m_\text{e}/m_\text{p})^3$ that were neglected in the current NAPT approach. Exact agreement with the results of the H2SPECTRE program suite \cite{SPECTRE2019} was obtained for the bound levels and the resonance positions were  
found in agreement with previously calculated collision energies \cite{Selg2012}, although the present values are more accurate by a factor of 30.
The resonance lifetimes are computed resulting in values ranging from 0.2~ms for X(7,21)$^*$ to 2~ns for X(11,13)$^*$.

\begin{figure}
\begin{center}
\includegraphics[width=1.0\linewidth]{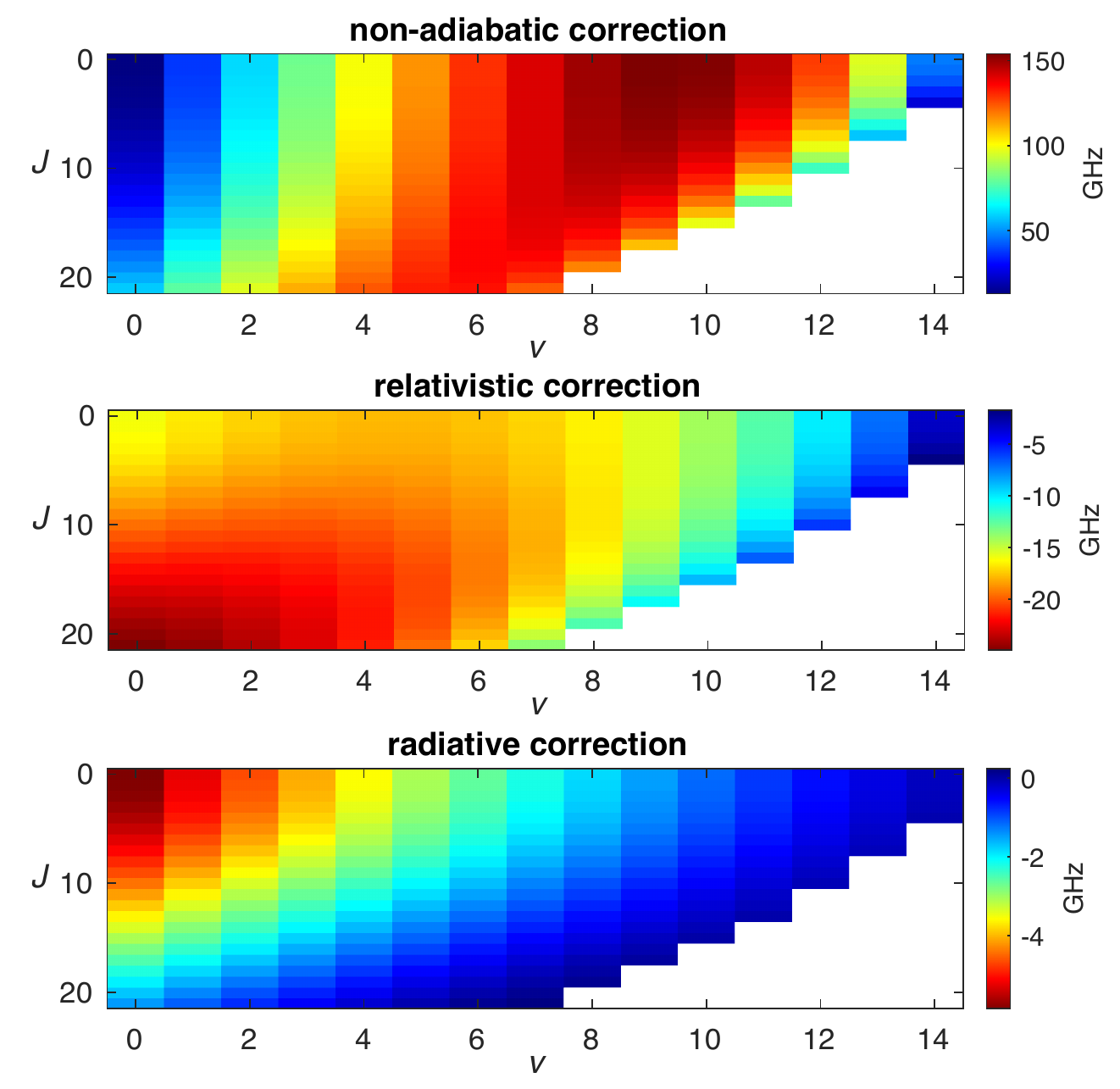}
\caption{\label{Heat}
Color-map displaying the magnitude of contributions to the energies of ($v,J$) levels in H$_2$, for both bound and quasi-bound levels, resulting from non-adiabatic, relativistic and radiative corrections to the BO-energies.
}
\end{center}
\end{figure}

Fig.~\ref{Heat} illustrates the non-adiabatic and leading order relativistic and radiative contributions to the H$_2$($v,J$) dissociation energies for bound states and for the observed shape resonances with $J=0-21$. It is evident, that nonadiabatic interactions affect mainly states with $v=7-11$, as these states possess the largest vibrational kinetic energy \cite{Van-Vleck1936}, with the maximum reached for $v=9$. The corrections of up to more than 100~GHz on the dissociation energies are positive, i.e., leading to larger dissociation energies, as it would be expected from interactions with distant states at higher energy. 
The calculations of the relativistic correction ($m_\text{e}\alpha^4$) on the order of several GHz reveal an increased sensitivity for levels with low $v$ and high $J$, as was tested in Ref.~\cite{Salumbides2011}. 
Computations for the leading-order QED-corrections $m\alpha^5$ (also depicted in Fig.~\ref{Heat}) show a larger sensitivity for levels with low $v$ and low $J$ and contribute on the order of 100~MHz to the resonance energies.

This suggests that narrow shape resonances are observed for a vibrational excitation that constitutes a precisely tailored test case for non-adiabatic effects in atomic collisions.
As an example, the position of the X(7,21)$^*$ resonance is shifted by ${\sim}10$~K, ${\sim}6$~K, ${\sim}0.7$~K and ${\sim}11$~mK by the adiabtic, non-adiabatic, relativistic and radiative correction, respectively (see also the inset of Fig.~\ref{ex_scheme}; it should be noted that the non-adiabatic correction results mainly from the $R$-dependent reduced mass). 

F0-level energies were obtained by solving the Schr\"{o}dinger equation using Born-Oppenheimer \cite{Silkowski2021}, adiabatic \cite{Wolniewicz1994} and relativistic \cite{Wolniewicz1998} potential energy curves, while including leading-order radiative corrections taken from Ref.~\cite{Korobov2018} for the hydrogen molecular ion and estimating the effects of nonadiabaticity based on Ref.~\cite{Yu1994}.
The computations for the F0($J$) level energies yield values accurate to $\sim 1$ \wn, which is sufficient to identify the lines in Fig.~\ref{Overview} and to prove the quasi-bound nature of the resonances.
The assignments are verified by the calculation of Franck-Condon factors (FCF) in the F-X system.
The shifting of the wave function of the H$_2^*$ shape resonances towards larger internuclear separation causes the favorable condition of enhanced FCF, giving increased intensities for those lines probing H$_2^*$, as is experimentally observed (see Fig.~\ref{Overview}).

\begin{figure}
\begin{center}
\includegraphics[width=1.0\linewidth]{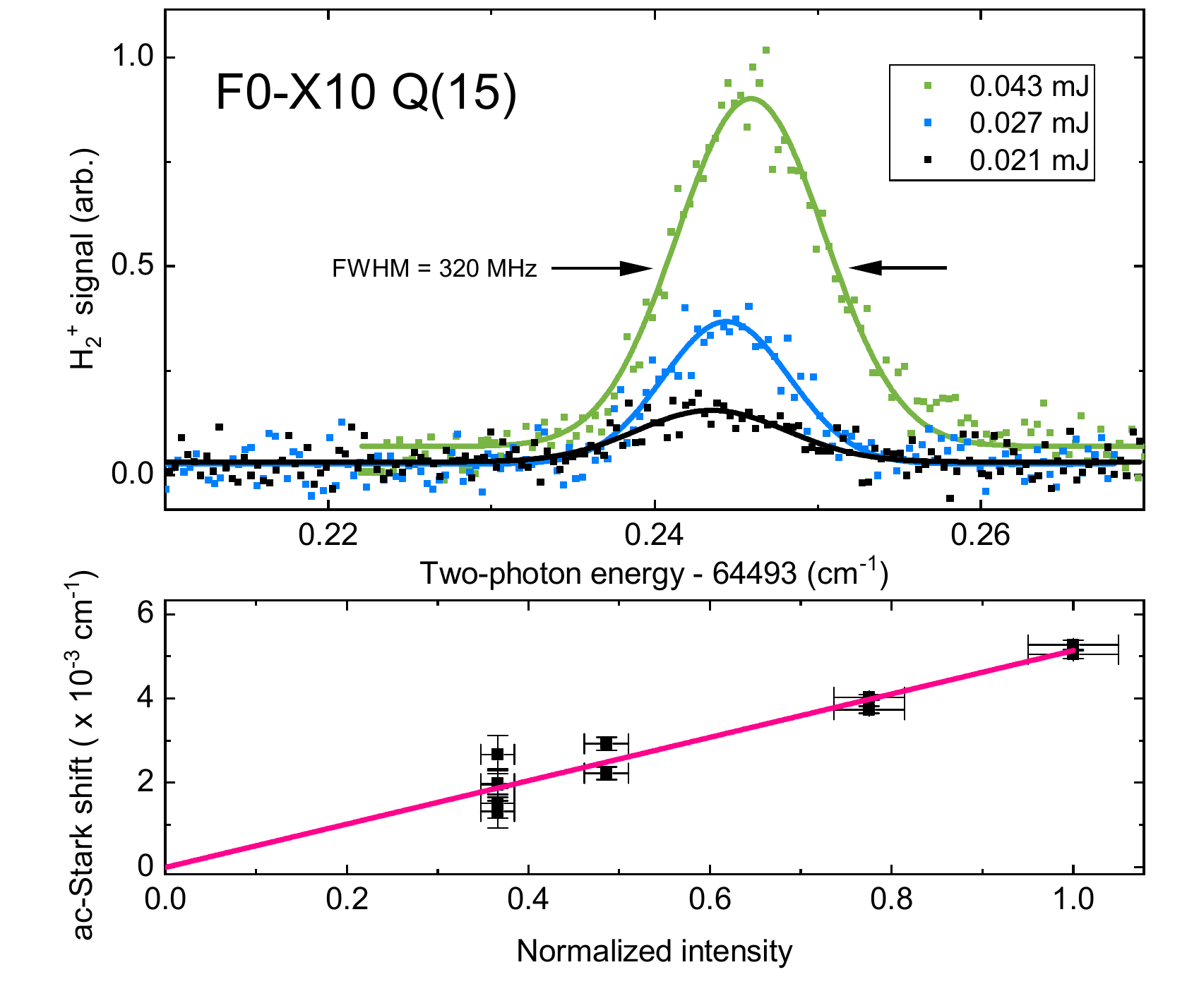}
\caption{\label{QB-spectra}
High-resolution excitation spectrum of a two-photon transition in the F-X system probing the shape resonance (10,15) via a Q-line.
The lower panel shows the AC-Stark extrapolation to deduce the field-free transition frequency.
}
\end{center}
\end{figure}

Subsequently the shape resonances $X(v,J)$: (7,21), (8,19), (9,17), (10,15) and (11,13) are probed in a  precision measurement, with the narrowband PDA-laser detecting the Doppler-free \F-\X\ two-photon  transition.
In Fig.~\ref{QB-spectra} an example of a line probing X(10,15) via a Q-line is presented.
These precision measurements are performed for varying intensities allowing for assessment of the AC-Stark effect in extrapolation to zero field as shown.
All five quasi-bound states have been excited via a Q-transition ($\Delta J=0$), while some are also probed via an S- or O-transitions.
The extrapolated zero-intensity values of the highly accurate transition frequencies are compiled in Table~\ref{tab:transition}. Also some transitions from bound states are subjected to a precision measurement.
Averaging over multiple measurement sequences yields an optimum uncertainty of 0.001 \wn, while it is 0.002 \wn\ for somewhat weaker transitions (30 and 60 MHz, respectively); for a detailed error budget see Supplemental Material. The uncertainty is larger for the short-lived H$_2^*$(11,13) resonance.

\begin{table}
\begin{threeparttable}
\caption{\label{tab:transition}
Measured frequencies for the two-photon F0-X transitions probing the quasi-bound levels H$_2^*$(X), with uncertainties indicated in parentheses. Also some transitions from bound H$_2$(X) levels are included.
}
\begin{tabular}{cclcl}
\hline
H$_2^*$(X)	&  Line & Exp. (\wn)    & Line & Exp. (\wn)    \\
\hline
 (7,21)* & Q(21) & 65\,441.3575 (9)    &   O(21) & 64\,979.430 (10)      \\
 (8,19)* & Q(19) & 65\,157.9413 (21)   &   O(19) & 64\,734.8098 (9) \\
 (8,19)* & S(19) & 65\,619.8599 (8)    &   & \\
 (9,17)* & Q(17) & 64\,837.2974 (19)   &   O(17) & 64\,454.775 (10)        \\
(10,15)* & Q(15) & 64\,493.2404 (9)    &   O(15) & 64\,152.970 (20) \\
(11,13)* & Q(13) & 64\,146.930 (20)    &  & \\
\hline
 (8,17)  & Q(17) & 66\,044.7046 (9)    &  &  \\
 (9,15)  & Q(15) & 65\,571.9063 (19)   & S(15) & 65\,954.4505 (10)     \\
\hline
\hline
\end{tabular}
\end{threeparttable}
\end{table}

\begin{table}[t]
\renewcommand{\arraystretch}{1.3}
\caption{Comparison of experimental and calculated collision energies $E_\text{Col}$ of observed shape resonances, and some bound states. Energy splittings with respect to X(10,15)$^*$ are determined from combination difference and shifted by the calculated dissociation energy of 186.4542(36)~\wn to give $E_\text{Col}$. Parentheses present the relative uncertainty to the X(10,15)$^*$ resonance. All values in \wn.
\label{tab:comb_diff}}
\begin{tabular}{ld{8}d{8}d{8}}
$(v,J)$ & \multicolumn{1}{c}{$E_\text{Col}^\text{exp}$} & \multicolumn{1}{c}{$E_\text{Col}^\text{calc}$}   & \multicolumn{1}{c}{$\Delta E_\text{Col}$}   \\
\hline
(7,21)$^*$  &   505.9314\,(28)    &   505.9310\,(9)    &   0.0004\,(29) \\
(8,19)$^*$  &   327.4290\,(25)    &   327.4291\,(7)    &   -0.0001\,(26) \\
(9,17)$^*$  &   224.9414\,(30)     &   224.9410(4)      &   0.0004\,(30) \\
(10,15)$^*$ &   186.4542     &   186.4542      &   \multicolumn{1}{c}{0}\\
(11,13)$^*$ &   192.495\,(28)      &   192.4945(6)       &   0.001\,(28) \\
\hline
(9,15)  &   -892.2117\,(21)  &   -892.2130(11)  &   0.0013\,(24) \\
(8,17)   &   -982.4658\,(25)     &    -982.4659\,(12)  &  0.0001\,(28)  \\
\hline
\end{tabular}
\end{table}

Although the transition frequencies probing the quasi-bound resonances were measured at high accuracy, this does not provide a direct means to determine collision energies.
In order to make a comparison with theory possible, combination differences between resonances and bound states have been computed with the results presented in Table~\ref{tab:comb_diff}. The X(10,15)$^*$ level was chosen as an anchor level and the collision energy was set to coincide with its theoretical value. Intervals can be calculated at higher accuracy than dissociation energies of single levels, because the neglected terms in the NAPT approach lead to an approximately equal shift of the X levels caused by the far distant states.   
Table~\ref{tab:comb_diff} demonstrates that the deviations between experimental and computed combination differences are well within 0.001 \wn\ (30~MHz).
This high level of agreement constitutes a test of the calculations of the potential energy curve, including all adiabatic, non-adiabatic, relativistic and QED contributions.
It tests the H$_2$ potential specifically at large internuclear distances and for the non-adiabatic corrections.
This importantly complements  previous precision tests that probe the bottom and deeper part of the H$_2$(X) potential well in experiments on vibrational splittings in H$_2$ \cite{Dickenson2013,Cheng2012,Kassi2014}
and on the dissociation energy $D_0$(H$_2$) \cite{Cheng2018,Holsch2019}.

As was noticed in Ref.~\cite{Londono2010}, a precise measurement of shape resonances can also be used to determine the scattering length.
We used the H$_2$ potential energy curve, now tested experimentally over a wide range of energies (including the continuum) and internuclear distances, to determine the singlet scattering length for the H(1s) + H(1s) collision. Because of the importance of non-adiabatic effects in this collision, different approaches for their treatment were vividly discussed in the literature \cite{Jamieson2010}. 
Whereas previously reported scattering lengths obtained by different authors within the BO and adiabatic approximations were found to agree, the values for the non-adiabatic scattering length varied between 0.3006~$a_0$ and 0.564~$a_0$ (see Table~3 in Ref.~\cite{Jamieson2010} and references therein), depending on the used reduced masses and effective correction potentials employed to account for the non-adiabatic interactions. 
Using the novel techniques presented here, which allow to include BO, adiabatic, non-adiabatic, relativistic and QED contributions up to $m_\text{e}\alpha^6$,  
we obtain a scattering length of $a = 0.2735^{39}_{31}~a_0$.

Non-adiabatic effects might also play a role in collisions of heavier atoms or in collisions of light atoms and molecules \cite{Paliwal2021}. In case no ab initio data is available, effects of this kind can be taken into account by using the atomic reduced mass when solving the radial Schr\"odinger equation. We found that such an approach leads to an error of 0.8\% and 0.001\% on the X(10,15)$^*$ and X(7,21)$^*$ resonance position, respectively. The scattering length found in this way deviates by 3\%. Although the scaling with the reduced mass results in nonadiabatic effects being reduced, for example by a factor of six in Li + Li collisions, the crude treatment of non-adiabatic effects using atomic reduced masses might lead to a very large disagreement given the experimental accuracy of typical cold atom experiments. As was pointed out in Ref.~\cite{kutzelnigg07a}, accurate $R$-dependent reduced masses accounting for the bulk of nonadiabatic effects can be even obtained from relatively crude molecular wavefunctions. This should allow testing the approach presented here for collisions involving heavier atoms with the currently reachable precision of ab initio calculations. Such precision studies uniquely straddle the interface between ultracold collisions in atomic physics and bond forming phenomena in chemistry.

Because of the long lifetime of the observed shape resonances, a measurement of their energy-level structure with kHz precision appears possible, which is interesting, given their low sensitivity to QED effects (see Fig.~3). This allows to specifically test the accuracy of relativistic four-body calculations, while avoiding the evaluation of QED effects, representing today the major source of theoretical uncertainty.

The authors thank Prof. Frederic Merkt (ETH Z\"urich) for discussions motivating this work. This work is financially supported by the European Research Council through an ERC Advanced grant (No: 670168). MB acknowledges NWO for a VENI grant (VI.Veni.202.140).

\pagebreak
\clearpage
\widetext
\section*{Supplemental Material}
The error budget for the frequency calibrations, presented in Table~\ref{tab:error}, contains a variety of contributions.
Minor contributions relate to the  calibration uncertainty of the cw-laser seeding the pulse-dye amplifier, of some 2 MHz (at the fundamental), while 
a residual Doppler shift from misalignment of the counter-propagating beams is reduced by Sagnac interferometry to below 3 MHz uncertainty. 
The chirp-induced frequency correction accounts for another 4.5 MHz uncertainty. 
The AC-Stark effect yields the largest contribution to the error budget. 
It is addressed by performing systematic measurements resulting in the AC-Stark slopes as indicated in Fig. \ref{QB-spectra}.
The uncertainty associated with AC-Stark results from the extrapolation to zero-power levels and depends on the obtained signal-to-noise ratio for individual lines.
In order to reduce the contributions of the AC-Stark effect various campaigns of remeasurement of the Stark-slopes were carried out, thus turning the systematic effect into a statistical distribution of results.
For some weak transition, for example F0-X11 Q(13) and F0-X10 O(15), this could not be done effectively, and only measurements at high laser power were performed leading to larger uncertainties.

\begin{table}[h!]
\renewcommand{\arraystretch}{1.2}
\caption{\label{tab:error}
Uncertainty budget for the measurements of two-photon F0-X transitions. The uncertainty for the AC-Stark extrapolations is estimated for individual transitions, where the total uncertainties are listed in Table~\ref{tab:transition}.
}
\begin{tabular}{lc}
Contribution & Uncertainty (MHz) \\
\hline
Lineshape fitting & 15 \\
Frequency calibration & 9 \\
CW-pulse offset (chirp)  & 18 \\
Residual Doppler effect & 3 \\
DC-Stark effect & $<$1\\
\hline
Subtotal (exl. AC-Stark) & 25\\
AC-Stark effect &  $3 - 60$\\
\end{tabular}
\end{table}

\end{document}